\documentclass[a4paper, twocolumn]{article}
\usepackage{graphicx}
\usepackage{amsmath}

\begin{document}

\title{The capacity of transmitting atomic qubit with light}
\author{Xiao-yu Chen \\
%EndAName
{\small {College of Information and Electronic Engineering, Zhejiang
Gongshang University, Hangzhou, 310018, China}}}
\date{}
\maketitle

\begin{abstract}
The quantum information transfer between a single photon and a
two-level atom is considered as a part of a quantum channel. The
channel is a degradable channel even when there are decays of the
atomic excited state and the single photon state, as far as the
total excitation of the combined initial state does not exceed 1.
The single letter formula for quantum capacity is obtained.
\newline
\end{abstract}

Recent experiment has realized the coherent transfer of coherent state of
light to and from the hyperfine states of an atom trapped within the modes
of high finesse optical cavity\cite{Boozer}. This may provide the basic
nodes of quantum networks. In the proposal for the implementation of quantum
networks\cite{Cirac}, each node is a quantum system that stores and locally
processes quantum information in quantum bits, atomic internal states with
long coherent times serve as these 'stationary' qubits, exchange of
information between the nodes of the network is realized by the transmission
of photons ('flying' qubits) over optical fiber. In the the experimental
setting, a '$\Lambda $' type three-level atom is used. A three-level system
can be reduced to a two-level system adiabatically for far detuning system
\cite{Cirac}. Thus a problem of basic interests is the interface of light
and two-level atom. We  will treat the quantum information transfer process
as a quantum channel (or a part of quantum channel) for the first time. The
whole process involves transfer of atomic state to field, the optical
transmission and transfer quantum state back to atom. The field is
restricted to single photon state and in each step of transfer the part that
receives quantum state is prepared as ground state or vacuum field state.
Field decay and atomic excited state decay are included in our model.

\textit{The state transfer: }The qubit is a two-level atom with ground state
$\left| \downarrow \right\rangle $ and excited state $\left| \uparrow
\right\rangle $. It interacts with a single-mode near-resonant cavity field
prepared in vacuum state $\left| 0\right\rangle $. The dynamics of the
system is given by the Jaynes-Cumming interaction Hamiltonian \cite{Jaynes} (%
$\hbar =1$)
\begin{equation}
\widehat{H}=\nu (a^{\dagger }a+\frac 12)+\frac \omega 2\sigma
_z+g(a^{\dagger }\sigma _{-}+a\sigma _{+}),
\end{equation}
where $\nu $ is the frequency of the field, $\omega $ is the atomic
transition frequency between the two levels, the coupling constant between
the atom and field is $g$, $a$ and $a^{\dagger }$ are the annihilation and
creation operators of the field, $\sigma _{-}$ and $\sigma _{+}$ are the
atomic state flip operators defined as $\sigma _{-}\left| \uparrow
\right\rangle =\left| \downarrow \right\rangle ,$ $\sigma _{-}\left|
\downarrow \right\rangle =0,\sigma _{+}\left| \downarrow \right\rangle
=\left| \uparrow \right\rangle ,$ $\sigma _{+}\left| \uparrow \right\rangle
=0,$ $\sigma _z=\sigma _{+}\sigma _{-}-\sigma _{-}\sigma _{+}$ and $\sigma
_z\left| \uparrow \right\rangle =\left| \uparrow \right\rangle ,\sigma
_z\left| \downarrow \right\rangle =-\left| \downarrow \right\rangle $. The
evolution operator of the system is
\begin{eqnarray}
\widehat{U} &=&\exp (-i\widehat{H}t)=e^{-i\nu \widehat{\xi }t}\{\cos (%
\widehat{\Omega }t) \\
&&-\frac i{\widehat{\Omega }}\sin (\widehat{\Omega }t)[\frac \Delta 2\sigma
_z+g(a^{\dagger }\sigma _{-}+a\sigma _{+})]\},
\end{eqnarray}
where $\widehat{\xi }=a^{\dagger }a+(1+\sigma _z)/2$ is the operator of the
total number of excitation of the system. $\Delta =\omega -\nu $ is the
detuning, and $\widehat{\Omega }=\sqrt{g^2\widehat{\xi }+\frac{\Delta ^2}4}.$
For initial states $\left| \downarrow 0\right\rangle $ or $\left| \uparrow
0\right\rangle $ , the evolved states are $\widehat{U}\left| \downarrow
0\right\rangle =e^{i\Delta t/2}\left| \downarrow 0\right\rangle ,\widehat{U}%
\left| \uparrow 0\right\rangle =e^{-i\nu t}\{[\cos \left( \Omega t\right)
-i\sin (\Omega t)\frac \Delta {2\Omega }]\left| \uparrow 0\right\rangle
-i\sin (\Omega t)\frac g\Omega \left| \downarrow 1\right\rangle \},$
respectively, where $\Omega =$ $\sqrt{g^2+\frac{\Delta ^2}4}.$ For a general
atomic density matrix $\rho _A=(1-p)\left| \downarrow \right\rangle
\left\langle \downarrow \right| $ $+r$ $\left| \downarrow \right\rangle
\left\langle \uparrow \right| $ $+r^{*}\left| \uparrow \right\rangle
\left\langle \downarrow \right| +p\left| \uparrow \right\rangle \left\langle
\uparrow \right| $, the evolved system state is $\widehat{U}\rho _A\otimes
\left| 0\right\rangle \left\langle 0\right| \widehat{U}^{\dagger }.$ At some
proper time $t,$ the atomic state is traced out, leaving the optical field
state $\rho _B=Tr_A(\widehat{U}\rho _A\otimes \left| 0\right\rangle
\left\langle 0\right| \widehat{U}^{\dagger }).$ In the photon number basis $%
\left| 0\right\rangle ,\left| 1\right\rangle $, it reads
\begin{equation}
\rho _B=\left[
\begin{array}{ll}
1-p\left| h_1(t)\right| ^2], & rh_1(t) \\
r^{*}h_1^{*}(t), & p\left| h_1(t)\right| ^2
\end{array}
\right] .
\end{equation}
with $h_1(t)=ie^{i(\Delta /2+\nu )t}\sin (\Omega t)\frac g\Omega .$

\textit{The conversion as a quantum channel: }The state of atomic system $A$
now is transferred to the state of optical field system $B$. If we neglect
the different natures of the atomic system and optical field for a while,
the process of the state transfer can be viewed as a quantum channel. It
maps a density matrix $\rho _A$ to another density matrix $\rho _B$. Denote
the conversion channel as $\mathcal{E},$ then
\begin{equation}
\rho _B=\mathcal{E}(\rho _A)=Tr_A(\widehat{U}\rho _A\otimes \left|
0\right\rangle \left\langle 0\right| \widehat{U}^{\dagger }).
\end{equation}
The channel $\mathcal{E}$ can be expressed with Kraus operator sum
representation with
\begin{eqnarray}
A_1 &=&e^{i\Delta t/2}\left| 0\right\rangle \left\langle \downarrow \right|
-ie^{-i\nu t}\sin (\Omega t)\frac g\Omega \left| 1\right\rangle \left\langle
\uparrow \right| , \\
A_2 &=&e^{-i\nu t}[\cos \left( \Omega t\right) -i\sin (\Omega t)\frac \Delta
{2\Omega }]\left| 0\right\rangle \left\langle \uparrow \right| ,
\end{eqnarray}
such that $\rho _B=A_1\rho _AA_1^{\dagger }+A_2\rho _AA_2^{\dagger }$ and $%
A_1^{\dagger }A_1+A_2^{\dagger }A_2=I.$

The procedure of quantum information transfer from atomic system to field
can be viewed as a quantum channel. Hence we can characterize the transfer
capability of the Jaynes-Cummings interaction with the quantum capacity of
the conversion channel if we are concerned with the quantum information
converted. The quantum capacity $Q$ measures the maximum amount of quantum
information that can be reliably transmitted (here transferred) though the
map $\mathcal{E}$ per channel use \cite{Lloyd}. The quantum capacity can be
computed by coherent information (CI) $I_c(\sigma ,\mathcal{E})=S(\mathcal{E}%
(\sigma ))-S(\sigma ^{QR^{\prime }})$ . Here $S(\varrho )=-$Tr$\varrho \log
_2\varrho $ is the von Neumann entropy, $\sigma $ is the input state, the
application of the channel $\mathcal{E}$ results the output state $\mathcal{E%
}(\sigma )$; $\sigma ^{QR^{\prime }}=$ $(\mathcal{E}\otimes \mathbf{I}%
)(\left| \psi \right\rangle \left\langle \psi \right| )$, with R referred to
the 'reference' system (the system under process is Q system, we denote $%
\sigma ^Q$ as $\sigma $ for simplicity), $\left| \psi \right\rangle $ is the
purification of the input state $\sigma $. The quantum channel capacity is
\begin{equation}
Q=\lim_{n\rightarrow \infty }\sup_{\sigma _n}\frac 1nI_c(\sigma _n,\mathcal{E%
}^{\otimes n}).
\end{equation}
This general result of capacity involves multiple use of the channel, it is
called the regulation of the channel. The coherent information is not a
convex function of the input state, hence it is very difficult to carry out
the capacity for a general channel map. Fortunately, if the channel map is
degradable, the capacity can be calculated with single letter formula \cite
{Devetak}
\begin{equation}
Q=\sup_\sigma I_c(\sigma ,\mathcal{E}).  \label{wave1}
\end{equation}
The degradability of a quantum channel is defined as follows: The sender $A$
prepares quantum state $\rho _A,$ the receiver $B$ obtains state $\rho _B=%
\mathcal{E}(\rho _A),$ where $\mathcal{E}$ is the channel map, the
environment $E$ obtains the state $\rho _E=\widetilde{\mathcal{E}}(\rho _A)$
in the transmission process, where $\widetilde{\mathcal{E}}$ is called
complementary channel. If all input state $\rho _A$, there exist a quantum
channel $\mathcal{N}$ such that
\begin{equation}
\mathcal{N(E}(\rho _A))=\widetilde{\mathcal{E}}(\rho _A),  \label{wave2}
\end{equation}
then the channel $\mathcal{E}$ is called degradable. The physical meaning of
the degradable condition is: the quantum state leaked to the environment can
be reconstructed by the receiver with some quantum map $\mathcal{T}$ .

In the following, we will prove that the conversion due to Jaynes-Cummings
model with proper interaction time is a degradable quantum channel for
vacuum initial field. The complementary channel $\widetilde{\mathcal{E}}$ of
$\mathcal{E}$ is defined by
\begin{equation}
\widetilde{\mathcal{E}}(\rho _A)=Tr_B(\widehat{U}\rho _A\otimes \left|
0\right\rangle \left\langle 0\right| \widehat{U}^{\dagger }),
\end{equation}
which is
\begin{equation}
\widetilde{\mathcal{E}}(\rho _A)=\left[
\begin{array}{ll}
1-p\left| h_2(t)\right| ^2, & rh_2(t) \\
r^{*}h_2^{*}(t), & p\left| h_2(t)\right| ^2
\end{array}
\right]
\end{equation}
in the basis $\left| \downarrow \right\rangle ,\left| \uparrow \right\rangle
,$ with $h_2(t)=e^{i(\Delta /2+\nu )t}[\cos \left( \Omega t\right) +i\sin
(\Omega t)\frac \Delta {2\Omega }].$ Note that $\left| h_1(t)\right|
^2+\left| h_2(t)\right| ^2=1.$ The quantum channel $\mathcal{N}$ then should
convert field to atomic system. It is implemented also by Jaynes-Cummings
interaction with Hamiltonian
\begin{equation}
\widehat{H}^{\prime }=\nu ^{\prime }(a^{\dagger }a+\frac 12)+\frac{\omega
^{\prime }}2\sigma _z+g^{\prime }(a^{\dagger }\sigma _{-}+a\sigma _{+}),
\end{equation}
then $\widehat{U^{\prime }}=\exp (-i\widehat{H^{\prime }}t^{\prime })$ and $%
\widehat{U}^{\prime }\left| \downarrow 0\right\rangle =e^{i\Delta ^{\prime
}t/^{\prime }2}\left| \downarrow 0\right\rangle ,\widehat{U}^{\prime }\left|
\downarrow 1\right\rangle =e^{-i\nu ^{\prime }t^{\prime }}\{[\cos \left(
\Omega ^{\prime }t^{\prime }\right) +i\sin (\Omega ^{\prime }t^{\prime })%
\frac{\Delta ^{\prime }}{2\Omega ^{\prime }}]\left| \downarrow
1\right\rangle -i\sin (\Omega ^{\prime }t^{\prime })\frac{g^{\prime }}{%
\Omega ^{\prime }}\left| \uparrow 0\right\rangle \}$ with $\Delta ^{\prime
}=\omega ^{\prime }-\nu ^{\prime }$ and $\Omega ^{\prime }=$ $\sqrt{%
g^{\prime 2}+\frac{\Delta ^{\prime 2}}4}.$ We have
\begin{equation}
\mathcal{N}(\rho _B)=Tr_B(\widehat{U}^{\prime }\left| \downarrow
\right\rangle \left\langle \downarrow \right| \otimes \rho _B\widehat{U}%
^{\prime \dagger }).
\end{equation}
We may set $\Delta ^{\prime }=0$ (resonant system), then
\begin{equation}
\mathcal{N}(\mathcal{E}(\rho _A))=\left[
\begin{array}{ll}
1-p\left| h_3(t)\right| ^2], & ih_3(t)r \\
ih_3^{*}(t)r^{*}, & p\left| h_3(t)\right| ^2
\end{array}
\right]
\end{equation}
with $h_3(t)=h_1(t)\sin (g^{\prime }t^{\prime }).$ The condition $\mathcal{N}%
(\mathcal{E}(\rho _A))=\widetilde{\mathcal{E}}(\rho _A)$ requires $1-\sin
^2(\Omega t)\frac{g^2}{\Omega ^2}=\sin ^2(\Omega t)\frac{g^2}{\Omega ^2}\sin
^2(g^{\prime }t^{\prime })$ and $\cos \left( \Omega t\right) +i\sin (\Omega
t)\frac \Delta {2\Omega }=e^{i\nu ^{\prime }t^{\prime }}\sin (\Omega t)\frac
g\Omega \sin (g^{\prime }t^{\prime }).$ The amplitude part of the second
equation is just the first equation, the phase factor can be adjusted by
properly choosing $\nu ^{\prime }.$ The condition reads
\begin{equation}
\left| \sin (\Omega t)\frac g\Omega \right| \geq \frac 1{\sqrt{2}}.
\end{equation}
With a similar calculation, it can be proved that the channel is
anti-degradable ($\mathcal{N}^{\prime }(\widetilde{\mathcal{E}}(\rho _A))=%
\mathcal{E}(\rho _A)$) for $\left| \sin (\Omega t)\frac g\Omega \right| \leq
\frac 1{\sqrt{2}},$ and the channel capacity is $0$ for anti-degradable
channel by no-cloning theorem.

The channel capacity then is
\begin{eqnarray}
Q &=&\max_{\rho _A}\{S(\mathcal{E}(\rho _A))-S[(\mathcal{E}\otimes \mathbf{I}%
)(\left| \psi \right\rangle \left\langle \psi \right| )]\}  \nonumber \\
&=&\max_{p\in [0,1]}\{H_2(\sin ^2(\Omega t)\frac{g^2}{\Omega ^2}p)  \nonumber
\\
&&-H_2((1-\sin ^2(\Omega t)\frac{g^2}{\Omega ^2})p)\},
\end{eqnarray}
with $H_2(x)=-x\log _2x-(1-x)\log _2(x)$ the binary entropy function, $%
\left| \psi \right\rangle $ is the purification of $\rho _A.$ $(\mathcal{E}%
\otimes \mathbf{I})(\left| \psi \right\rangle \left\langle \psi \right| )$
is a $4\times 4$ matrix of rank $2.$ The two nonzero eigenvalues can easily
be obtained. The explicit expression of the matrix for amplitude damping
channel was given in the appendix of Ref. \cite{Giovannetti}. With a few
modification, the matrix $(\mathcal{E}\otimes \mathbf{I})(\left| \psi
\right\rangle \left\langle \psi \right| )$ can be obtained.

\textit{Transmission of atomic qubit with light: }The total procedure of
quantum information conversion would not be ceased at the stage of the
transfer of atomic to field system. The next two steps are optical field
state transmission over fiber and convert the photonic state back to atomic
state. Typically, optical fiber will damp the state with loss, and the
inverse conversion may not be ideal. The whole process of transmitting
atomic state with light consists at least three parts: (1) atom to field
conversion, (2) transmission on fiber (or free space), (3) the inverse
conversion. We may consider these three steps as quantum channel maps which
are denoted with $\mathcal{E},\mathcal{T},\mathcal{E}^{\prime }$,
respectively. The optical transmission channel is characterized by the
transmittance $T.$ The inverse conversion $\mathcal{E}^{\prime }$is just the
channel map $\mathcal{N}$ with the free choice of all its parameters. The
whole process then is represented by a concatenate channel of $\mathcal{C=E}%
^{\prime }\circ \mathcal{T}\circ \mathcal{E}$. The quantum systems are
denoted as $A,B,C,D$ for the sender, the field before transmission, the
field after transmission, the receiver, respectively. We have $\rho _B=%
\mathcal{E}(\rho _A),$ $\rho _C=\mathcal{T}(\rho _B),\rho _D=\mathcal{E}%
^{\prime }(\rho _C).$ The atomic state sent is $\rho =\rho _A,$ the atomic
state received is $\rho ^{\prime }=\rho _D.$ Note that the lossy fiber $%
\mathcal{T}$ is a degradable channel \cite{Caruso}, and we have proved that
the channel of atomic to field conversion $\mathcal{E}$ is a degradable
channel. It can be proved that the inverse conversion is also degradable. So
we can anticipate that the concatenate channel $\mathcal{C}$ is also
degradable. In the following, we will prove that it is really the case.

The action of the lossy fiber is equivalent to that of a beam splitter. The
channel $\mathcal{T}$ can be represented by beam splitter operator. Now the
optical state is in the Hilbert space with basis $\left| 0\right\rangle $
and $\left| 1\right\rangle ,$ the action of the channel can be simplified to
an operator $\widehat{V}$ with
\begin{eqnarray}
\widehat{V}\left| 0\right\rangle _B\left| 0\right\rangle _C &=&\left|
0\right\rangle _B\left| 0\right\rangle _C, \\
\widehat{V}\left| 1\right\rangle _B\left| 0\right\rangle _C &=&\sqrt{1-T}%
\left| 1\right\rangle _B\left| 0\right\rangle _C+\sqrt{T}\left|
0\right\rangle _B\left| 1\right\rangle _C.
\end{eqnarray}
The concatenate channel $\mathcal{C}$ maps the input state $\rho $ to $\rho
^{\prime }=\mathcal{C}(\rho )$, and $\mathcal{C}(\rho )=Tr_{ABC}[\widehat{U}%
^{\prime }\widehat{V}\widehat{U}\rho \otimes \left| 0\right\rangle
_B\left\langle 0\right| \otimes \left| 0\right\rangle _C\left\langle
0\right| \otimes \left| \downarrow \right\rangle _D\left\langle \downarrow
\right| \widehat{U}^{\dagger }\widehat{V}^{\dagger }\widehat{U}^{\prime
\dagger }].$ The detail calculation gives
\begin{equation}
\mathcal{C}(\rho )=\left[
\begin{array}{ll}
1-p\left| h_4(t,t^{\prime })\right| ^2, & rh_4(t,t^{\prime }) \\
r^{*}h_4(t,t^{\prime })^{*} & p\left| h_4(t,t^{\prime })\right| ^2
\end{array}
\right]
\end{equation}
in the atomic basis $\left| \downarrow \right\rangle $ and $\left| \uparrow
\right\rangle ,$ where $h_4(t,t^{\prime })=-e^{i(\Delta /2+\nu )t+i(\Delta
^{\prime }/2+\nu ^{\prime }t^{\prime })}\sin (\Omega t)\frac g\Omega \sqrt{T}%
\sin (\Omega ^{\prime }t^{\prime })\frac{g^{\prime }}{\Omega ^{\prime }}.$

We have three kinds of complementary channels $\widetilde{\mathcal{E}}$ , $%
\widetilde{\mathcal{T}}\circ \mathcal{E}$ and $\widetilde{\mathcal{E}%
^{\prime }}\circ $ $\mathcal{T}\circ \mathcal{E}$. They map the input state $%
\rho $ to states that leak to environment at the three steps. The
degradability of $\mathcal{C}$ requires the existence of channel maps $%
\mathcal{N}_1,\mathcal{N}_2,\mathcal{N}_3$ such that $\mathcal{N}_1\circ
\mathcal{C=}\widetilde{\mathcal{E}},$ $\mathcal{N}_2\circ \mathcal{C}=%
\widetilde{\mathcal{T}}\circ \mathcal{E},$ $\mathcal{N}_3\circ \mathcal{C}=%
\widetilde{\mathcal{E}^{\prime }}\circ $ $\mathcal{T}\circ \mathcal{E}$ for
all input state $\rho .$ The channels $\mathcal{N}_i$ can be constructed as
in the previous section. The conditions of degradability are $1-\sin
^2(\Omega t)\frac{g^2}{\Omega ^2}\leq \left| h_4(t,t^{\prime })\right|
^2,1-T\sin ^2(\Omega t)\frac{g^2}{\Omega ^2}\leq \left| h_4(t,t^{\prime
})\right| ^2$and $1-\left| h_4(t,t^{\prime })\right| ^2\leq \left|
h_4(t,t^{\prime })\right| ^2,$respectively. The three conditions can be
combined to the condition $1-\left| h_4(t,t^{\prime })\right| ^2\leq \left|
h_4(t,t^{\prime })\right| ^2,$ which is
\begin{equation}
\left| \sin (\Omega t)\frac g\Omega \sqrt{T}\sin (\Omega ^{\prime }t^{\prime
})\frac{g^{\prime }}{\Omega ^{\prime }}\right| \geq \frac 1{\sqrt{2}}.
\end{equation}
The channel capacity is
\begin{equation}
Q=\max_{p\in [0,1]}\{H_2(p\left| h_4(t,t^{\prime })\right|
^2)-H_2(p(1-\left| h_4(t,t^{\prime })\right| ^2)\}
\end{equation}
for $\left| h_4(t,t^{\prime })\right| ^2\geq \frac 12$ and $Q=0$ for $\left|
h_4(t,t^{\prime })\right| ^2<\frac 12.$

\textit{The effect of decay: }To comply with experimental environment, we
introduce photonic decay rate $k$ and atomic excited decay rate $\gamma $
into our system. For a two-level system reduced from three-level $\Lambda $
cavity trapped atom, the two levels are two hyperfine ground states. The
decay rate $\gamma $ can be omitted comparing with other parameters, while
photonic decay rate $k$ should be considered in order to read out the
quantum state encoded in the atom. The master equation of the whole atom and
field system is

\begin{equation}
\frac{d\rho }{dt}=-i[\widehat{H},\rho ]+\mathcal{L}_1\rho +\mathcal{L}_2\rho
.
\end{equation}
with $\mathcal{L}_1\rho =\frac k2(2a\rho a^{\dagger }-a^{\dagger }a\rho
-\rho a^{\dagger }a),$ $\mathcal{L}_2\rho =\frac \gamma 2(2\sigma _{-}\rho
\sigma _{+}-\sigma _{+}\sigma _{-}\rho -\rho \sigma _{+}\sigma _{-}).$ The
basis of the whole system density matrix $\rho $ are $\left| \downarrow
0\right\rangle ,\left| \downarrow 1\right\rangle ,\left| \uparrow
0\right\rangle ,\left| \uparrow 1\right\rangle .$ To simplify the notation
of the entries of $\rho ,$ we abbreviate the basis as $\left| m\right\rangle
$ with $m=0,1,2,3$ corresponding to $\left| \downarrow 0\right\rangle
,\left| \downarrow 1\right\rangle ,\left| \uparrow 0\right\rangle ,\left|
\uparrow 1\right\rangle ,$ respectively. Then we have the equations for $%
\rho _{mn}.$ In our system, the total excitation of the initial state is
assumed to be $1.$ In the later evolution, this number of the total
excitation can not exceed $1.$ $\widehat{H}$ is a excitation number
conserved Hamiltonian, the decay can only decrease the excitation number.
Hence, in such an initial condition, $\rho _{m3}(t)=\rho _{3m}(t)\equiv 0$
for all $m$. We here consider the transfer of single photonic state to atom
(In the situation of total excitation limited to $1$, there is the symmetry
between the field and atom, thus the case of transfer of quantum information
from atom to field can be obtained accordingly). Suppose the general initial
state of $\left| \downarrow \right\rangle $ $\left\langle \downarrow \right|
\otimes \rho _{photon}$ with $\rho _{photon}=\overline{p}\left|
0\right\rangle \left\langle 0\right| +r$ $\left| 0\right\rangle \left\langle
1\right| +r^{*}\left| 1\right\rangle \left\langle 0\right| +p\left|
1\right\rangle \left\langle 1\right| $ and $\overline{p}=1-p$ , then $\rho
_{00}(0)=\overline{p},\rho _{02}(0)=r,\rho _{20}(0)=r^{*},\rho _{22}(0)=p,$
and all other entries are $0$ initially. The linear equations of $\rho _{mn}$
($m,n=0,1,2$) can be solved with Laplacian transformations. The solutions
are $\rho _{00}(t)=1-\rho _{11}(t)-\rho _{22}(t),$ and
\begin{eqnarray}
\rho _{01}(t) &=&\frac{igr}{X+iY}e^{-\frac 12k_1t+i(\nu +\frac \Delta 2)t}
\nonumber \\
&&[e^{\frac 12(X+iY)}-e^{-\frac 12(X+iY)}],
\end{eqnarray}
\begin{eqnarray}
\rho _{02}(t) &=&\frac r{2(X+iY)}e^{-\frac 12k_1t+i(\nu +\frac \Delta 2)t}
\nonumber \\
&&\{(k_2+i\Delta )[e^{\frac 12(X+iY)}-e^{-\frac 12(X+iY)}]  \nonumber \\
&&+(X+iY)[e^{\frac 12(X+iY)}+e^{-\frac 12(X+iY)}]\},
\end{eqnarray}
\begin{equation}
\rho _{11}(t)=2pg^2\eta (\cosh Xt-\cos Yt),
\end{equation}

\begin{eqnarray}
\rho _{12}(t) &=&pg\eta [(\Delta -ik_2)(\cosh Xt-\cos Yt)  \nonumber \\
&&+(Y-iX)(\sinh Xt-i\sin Yt),
\end{eqnarray}
\begin{eqnarray}
\rho _{22}(t) &=&p\eta [(X^2+\Delta ^2+2g^2)\cosh Xt  \nonumber \\
&&+(k_2X+\Delta Y)\sinh Xt  \nonumber \\
&&+(Y^2-\Delta ^2-2g^2)\cos Yt  \nonumber \\
&&+(k_2Y-\Delta X)\sin Yt].
\end{eqnarray}
where $k_{1,2}=(k\pm \gamma )/2,\eta =\frac{e^{-k_1t}}{X^2+Y^2},$ and $X=%
\sqrt{\frac 12(\sqrt{z^2+4k_2^2\Delta ^2}-z)},Y=\sqrt{\frac 12(\sqrt{%
z^2+4k_2^2\Delta ^2}-z)},$ with $z=4g^2+\Delta ^2-k_2^2.$ We now consider
the process of quantum state transfer from field to atom as a quantum
channel map $\mathcal{E}$. Then by tracing out the field freedom of $\rho
(t),$ we obtain $\mathcal{E}(\rho _{photon})=[1-\rho _{11}(t)]\left|
\downarrow \right\rangle \left\langle \downarrow \right| +\rho
_{01}(t)]\left| \downarrow \right\rangle \left\langle \uparrow \right| $ $%
+\rho _{10}(t)]\left| \uparrow \right\rangle \left\langle \downarrow \right|
+\rho _{11}(t)]\left| \uparrow \right\rangle \left\langle \uparrow \right| .$
By tracing out the atomic freedom of $\rho (t),$ we obtain the complimentary
channel map $\widetilde{\mathcal{E}}(\rho _{photon})=[1-\rho _{22}(t)]\left|
0\right\rangle \left\langle 0\right| +\rho _{02}(t)]\left| 0\right\rangle
\left\langle 1\right| $ $+\rho _{20}(t)]\left| 1\right\rangle \left\langle
0\right| +\rho _{22}(t)]\left| 1\right\rangle \left\langle 1\right| .$ It is
not difficult to prove that $\left| \rho _{01}(t)\right| ^2p=\rho
_{11}\left| \gamma \right| ^2,$and $\left| \rho _{02}(t)\right| ^2p=\rho
_{22}\left| \gamma \right| ^2,$ thus, the maps of the channel and the
complimentary channel are
\begin{eqnarray}
\mathcal{E}(\rho _{photon}) &=&\left[
\begin{array}{ll}
1-p\left| h_5(t)\right| ^2 & \gamma h_5(t) \\
\gamma ^{*}h_5^{*}(t) & p\left| h_5(t)\right| ^2
\end{array}
\right] , \\
\widetilde{\mathcal{E}}(\rho _{photon}) &=&\left[
\begin{array}{ll}
1-p\left| h_6(t)\right| ^2 & \gamma h_6(t) \\
\gamma ^{*}h_6^{*}(t) & p\left| h_6(t)\right| ^2
\end{array}
\right]
\end{eqnarray}
in their own basis. Where $h_5(t)=\sqrt{2g^2\eta (\cosh Xt-\cos Yt)}%
e^{i\theta _1(t)},h_6(t)=\sqrt{\eta }[(X^2+\Delta ^2+2g^2)\cosh Xt$ $%
+(k_2X+\Delta Y)\sinh Xt$ $+(Y^2-\Delta ^2-2g^2)\cos Yt$ $+(k_2Y-\Delta
X)\sin Yt]^{\frac 12}e^{i\theta _2(t)}$ for some phase factors $\theta _1(t)$
and $\theta _2(t).$ The degradability condition is $\left| h_5(t)\right|
\geq \left| h_6(t)\right| $ , which can be written as
\begin{eqnarray}
(X^2+\Delta ^2)\cosh Xt+(k_2X+\Delta Y)\sinh Xt &&  \nonumber \\
+(Y^2-\Delta ^2)\cos Yt+(k_2Y-\Delta X)\sin Yt] &\leq &0
\end{eqnarray}
When the channel is degradable, the capacity is
\begin{equation}
Q=\max_{p\in [0,1]}\{H_2(p\left| h_5(t)\right| ^2)-H_2(p(1-\left|
h_5(t)\right| ^2)\},
\end{equation}
otherwise $Q=0.$

\textit{Conclusion: }The quantum capacity of the conversion
channel of quantum information between single photon number state
and two-level atomic state is obtained as a function of coupling
rate, detuning, decay rates and the operation time. For an
efficient quantum information transfer, a proper operation time
should be chosen and the coupling rate should be high enough
comparing with the detunging and decay rates.\\

This work is supported by National Natural Science Foundation of
China under Grant No. 10575092.

\end{document}